\documentclass[aps,prl,twocolumn,showpacs,superscriptaddress]{revtex4}

\usepackage{graphics}
\usepackage{epsfig}
\usepackage{epstopdf}
\usepackage{float}

\usepackage[T1]{fontenc}
\usepackage{mathptmx}

\begin{document}

%@twocolumnfalse\endcsname

\title{Prediction of fully metallic $\sigma$-bonded boron framework induced high superconductivity \protect\\ above 100 K in thermodynamically stable Sr$_{2}$B$_{5}$ at 40 GPa}

\author{Xin Yang}
\affiliation{Key Laboratory of Material Simulation Methods \& Software of Ministry of Education, College of Physics, Jilin University, Changchun 130012, China}  
\affiliation{State Key Laboratory for Superhard Materials, College of Physics, Jilin University, Changchun 130012, China}
\author{Wenbo Zhao}
\affiliation{Key Laboratory of Material Simulation Methods \& Software of Ministry of Education, College of Physics, Jilin University, Changchun 130012, China}  
\affiliation{State Key Laboratory for Superhard Materials, College of Physics, Jilin University, Changchun 130012, China}
\affiliation{International Center of Future Science, Jilin University, Changchun 130012, China}
\author{Liang Ma}
\affiliation{Key Laboratory of Material Simulation Methods \& Software of Ministry of Education, College of Physics, Jilin University, Changchun 130012, China}  
\affiliation{State Key Laboratory for Superhard Materials, College of Physics, Jilin University, Changchun 130012, China}
\affiliation{International Center of Future Science, Jilin University, Changchun 130012, China}
\author{Wencheng Lu}
\affiliation{Key Laboratory of Material Simulation Methods \& Software of Ministry of Education, College of Physics, Jilin University, Changchun 130012, China}  
\author{Xin Zhong}
\affiliation{Key Laboratory of Material Simulation Methods \& Software of Ministry of Education, College of Physics, Jilin University, Changchun 130012, China}  
\author{Yu Xie}
\email{xieyu@jlu.edu.cn}
\affiliation{Key Laboratory of Material Simulation Methods \& Software of Ministry of Education, College of Physics, Jilin University, Changchun 130012, China}  
\affiliation{Key Laboratory of Physics and Technology for Advanced Batteries of Ministry of Education, College of Physics, Jilin University, Changchun 130012, China}
\author{Hanyu Liu}
\email{hanyuliu@jlu.edu.cn}
\affiliation{Key Laboratory of Material Simulation Methods \& Software of Ministry of Education, College of Physics, Jilin University, Changchun 130012, China}  
\affiliation{State Key Laboratory for Superhard Materials, College of Physics, Jilin University, Changchun 130012, China}
\affiliation{International Center of Future Science, Jilin University, Changchun 130012, China}
\affiliation{Key Laboratory of Physics and Technology for Advanced Batteries of Ministry of Education, College of Physics, Jilin University, Changchun 130012, China}
\author{Yanming Ma}
\email{mym@jlu.edu.cn}
\affiliation{Key Laboratory of Material Simulation Methods \& Software of Ministry of Education, College of Physics, Jilin University, Changchun 130012, China}  
\affiliation{State Key Laboratory for Superhard Materials, College of Physics, Jilin University, Changchun 130012, China}
\affiliation{International Center of Future Science, Jilin University, Changchun 130012, China}
\date{\today}

\begin{abstract}
Metal borides have been considered as potential high-temperature superconductors since the discovery of record-holding 39 K superconductivity in bulk MgB$_{2}$. In this work, we identified a superconducting yet thermodynamically stable $F$$\overline{4}$3$m$ Sr$_{2}$B$_{5}$ at 40 GPa with a unique covalent $sp^{3}$-hybridized boron framework through extensive first-principles structure searches. Remarkably, solving the anisotropic Migdal-Eliashberg equations resulted in a high superconducting critical temperature ($T_\mathrm{c}$) around 100 K, exceeding the boiling point (77 K) of liquid nitrogen. Our in-depth analysis revealed that the high-temperature superconductivity mainly originates from the strong coupling between the metalized $\sigma$-bonded electronic bands and $E$ phonon modes of boron atoms. Moreover, anharmonic phonon simulations suggest that $F$$\overline{4}$3$m$ Sr$_{2}$B$_{5}$ might be recovered to ambient pressure. Our current findings provide a prototype structure with a full $\sigma$-bonded boron framework for the design of high-$T_\mathrm{c}$ superconducting borides that may expand to a broader variety of lightweight compounds.

\end{abstract}

 
\maketitle

The quest for high-temperature superconductors has been the long-sought target since the discovery of superconductivity in solid mercury in 1911 \cite{1911}. According to the Bardeen-Cooper-Schrieffer (BCS) theory of superconductivity, light-element compounds are considered as promising candidates for high superconducting critical temperature ($T_\mathrm{c}$) superconductors since they possess high Debye temperature, which is proportional to the $T_\mathrm{c}$ \cite{suhl1959bardeen}. Recently, the theory-orientated experiments have established near-room-temperature superconductivity in a class of hydrides at high pressure, such as H$_{3}$S, LaH$_{10}$, and CaH$_{6}$ \cite{drozdov2015conventional,peng2017hydrogen,drozdov2019superconductivity,somayazulu2019evidence,ma2022high,li2014metallization,liu2017potential,wang2012superconductive}. Nevertheless, the pressure required to realize these superconducting hydrides is exceptionally high and challenging to achieve. Therefore, it is of significant importance to search for alternative high-$T_\mathrm{c}$ superconductors at moderate pressures and even at ambient pressure for practical applications.

At low pressure, MgB$_{2}$ is the BCS superconductor with the highest $T_\mathrm{c}$ of 39 K owing to its high Debye temperature and strong electron-phonon coupling (EPC), primarily originating from the metalized $\sigma$-bonding bands of in-plane $sp^{2}$-hybridized boron atoms \cite{ nagamatsu2001superconductivity,kortus2001superconductivity}. Thus, metal borides characterized by metallic $\sigma$ bonds have attracted considerable attention in searching for high-$T_\mathrm{c}$ superconductors. Besides MgB$_{2}$, a class of MB$_{2}$, MB$_{6}$, and MB$_{12}$ (M represents transition metals) have also been experimentally identified to be superconducting with $T_\mathrm{c}$ below 20 K at ambient pressure \cite{fisk1971superconducting,schneider1987electron,lortz2006superconductivity,akopov2017rediscovering}. Recently, Pei \textit{et al}. discovered that $\alpha$-MoB$_{2}$, isostructural with MgB$_{2}$, exhibits the second-highest $T_\mathrm{c}$ of 32 K at 100 GPa among all known borides \cite{pei2023pressure}. However, as the main contribution at the Fermi level of these borides is not derived from $\sigma$ bonds of boron atoms, none of them have achieved a $T_\mathrm{c}$ surpassing that of MgB$_{2}$ yet.

In addition to experiments, much effort has also been devoted to theoretical investigations to design borides with high $T_\mathrm{c}$. On one hand, thermodynamically stable boride superconductors, such as BeB$_{6}$, CaB, and SrB characterized by deformed $sp^{2}$-hybridization \cite{wu2016coexistence,cui2022superconducting,shah2013stability}, exhibit lower $T_\mathrm{c}$ compared to MgB$_{2}$ due to the weaker metallic $\sigma$ bonds, which is similar to the experimental observation. On the other hand, a series of MB$_{2}$ monolayer and MB$_{4}$ trilayers, such as AlB$_{2}$, MgB$_{4}$, and InB$_{4}$ \cite{zhao2019two,zhao2020mgb,wang2021three}, have been predicted to exhibit high $T_\mathrm{c}$ due to the strong metallic $\sigma$ bonds originating from $sp^{2}$-hybridized boron atoms, while their producibility is unclear since their thermodynamic stability remains unknown. Similarly, various metastable $sp^{3}$-hybridized borocarbides and boronitrides have also been predicted to exhibit higher $T_\mathrm{c}$ than MgB$_{2}$ \cite{geng2023conventional,gai2022van,ding2022ambient,zhu2020carbon,zhang2022path,wang2021high}, although none of them has been experimentally synthesized to date. Thus, it is encouraging to design metal borides with favorable thermodynamic stability and strong metallic $sp^{2}$/$sp^{3}$ $\sigma$-bonding characteristics that may lead to synthesizable high-temperature superconductors with a $T_\mathrm{c}$ higher than MgB$_{2}$.

In this work, we performed a comprehensive first-principles structure searches for the alkaline-earth (AE) metal borides (M$_{x}$B$_{y}$, M = Mg, Ca, Sr, Ba, $x$ = 1-3, $y$ = 1-8) at pressures below 100 GPa via the swarm intelligence-based CALYPSO method \cite{wang2010crystal,wang2012calypso}. Besides the reproduced already known phases \cite{esfahani2017novel,shah2013stability,kolmogorov2012pressure,cui2022superconducting}, we predicted several new thermodynamically stable borides under high pressures. Among them, the $F$$\overline{4}$3$m$ Sr$_{2}$B$_{5}$ consists of a unique fully $sp^{3}$-hybridized $\sigma$-bonded boron framework, where the density of states at Fermi level is mainly contributed by $\sigma$-bond. This results in a high $T_\mathrm{c}$  above 100 K, not only surpassing the boiling temperature (77 K) of liquid nitrogen but also setting a new record for the highest $T_\mathrm{c}$ among all known thermodynamically stable boride superconductors.

\begin{figure}[!t]
	\begin{center}
		\epsfxsize=8.5cm
		\epsffile{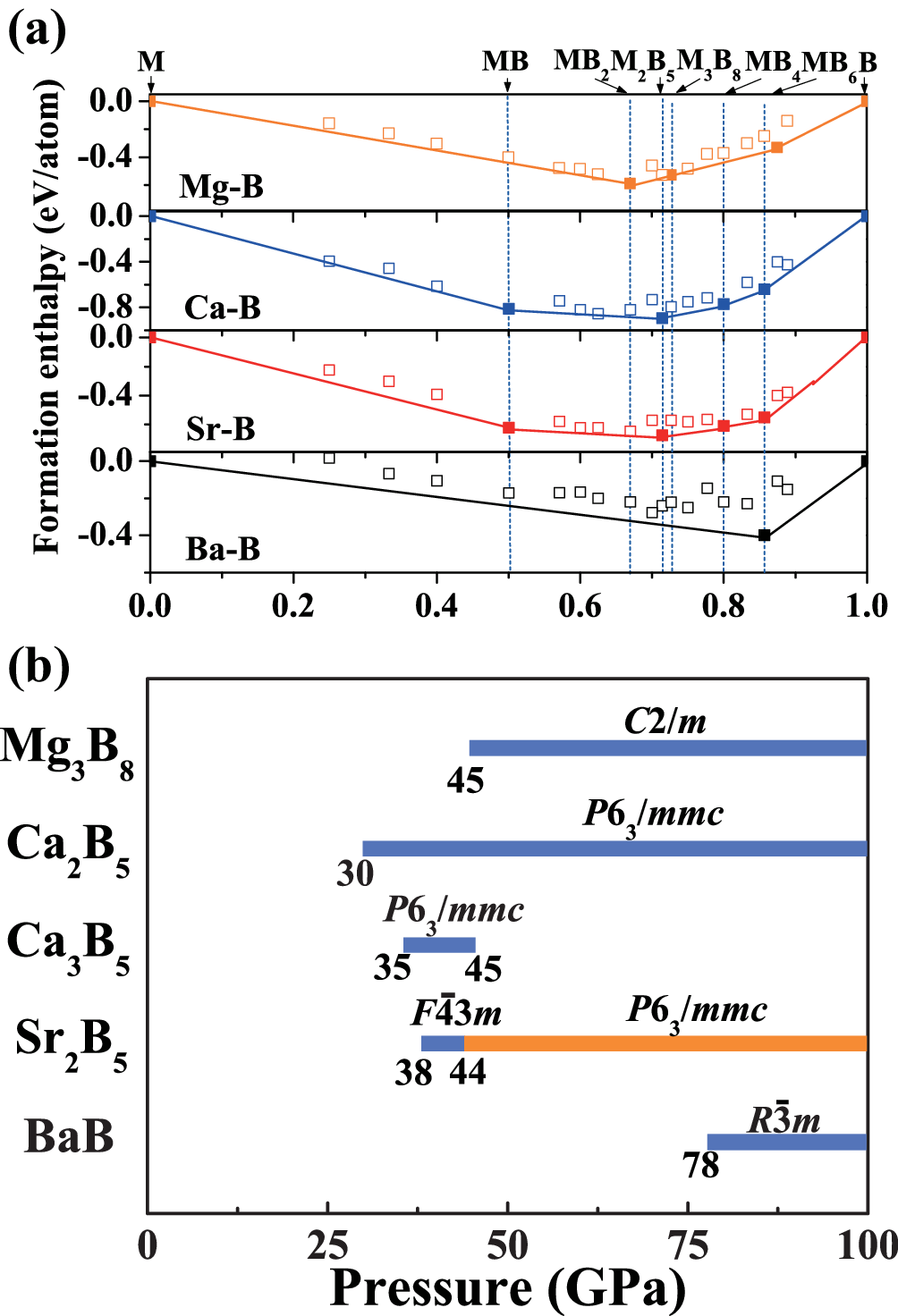}
	\end{center}
	\caption{(Color online) Phase stabilities of various AE borides M$_{x}$B$_{y}$ (M = Mg, Ca, Sr, Ba, $x$ = 1-3, $y$ = 1-8) (a) Convex hull diagrams for M$_{x}$B$_{y}$ at 50 GPa. (b) Pressure-composition phase diagram of newly-predicted Mg$_{3}$B$_{8}$, Ca$_{2}$B$_{5}$, Ca$_{3}$B$_{5}$, Sr$_{2}$B$_{5}$, and BaB. }
	\label{fig:phase}
\end{figure}

\begin{figure}[!t]
	\begin{center}
		\epsfxsize=8.5cm
		\epsffile{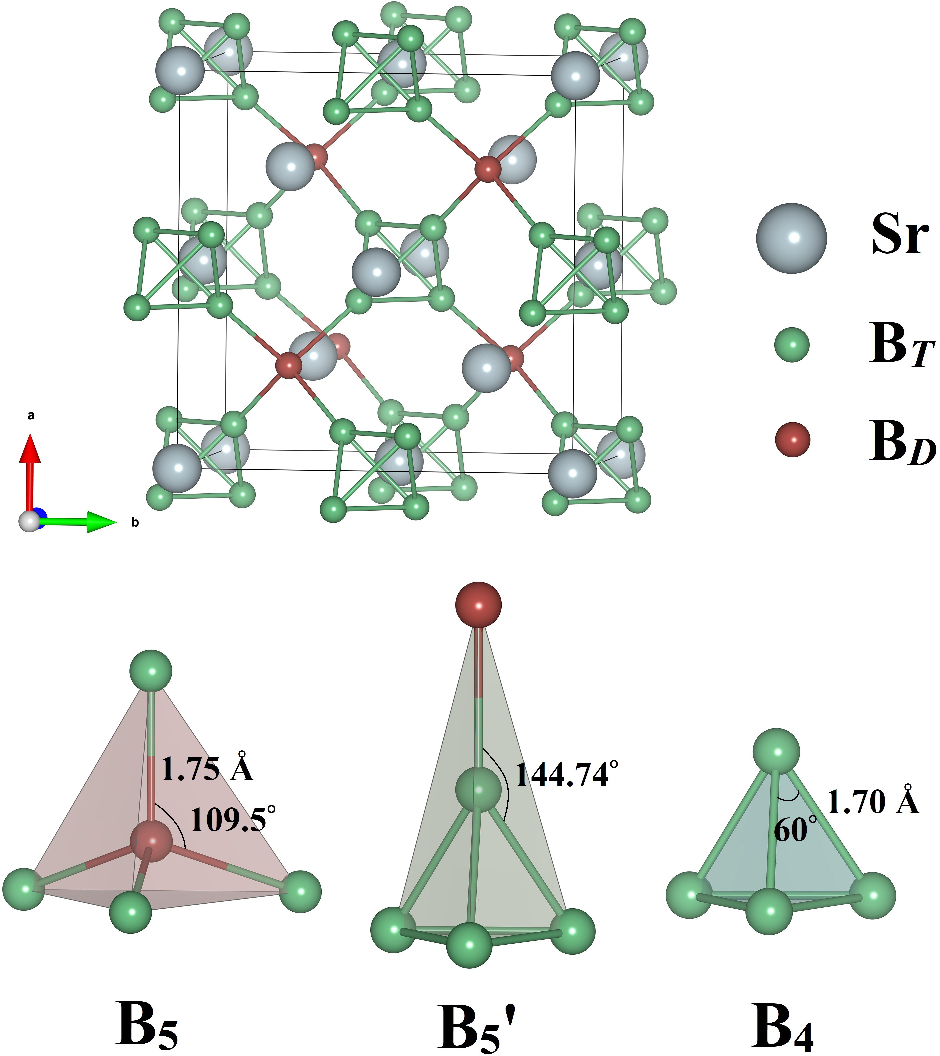}
	\end{center}
	\caption{(Color online) The side view of the structure of $F$$\overline{4}$3$m$ Sr$_{2}$B$_{5}$ at 40 GPa. The large and small spheres represent the Sr and B atoms, respectively. }
	\label{fig:str}
\end{figure}

Our main structure searching results of AE-metal borides M$_{x}$B$_{y}$ below 100 GPa are presented in the convex hull and phase diagrams of Fig. \ref{fig:phase} and Fig. S1 \cite{SI}.
 Considering the important contribution of zero-point energy (ZPE) to the free energy of compounds containing light elements \cite{hu2013pressure,peng2012predicted}, the formation energies of the metal borides were calculated with the inclusion of ZPE. %The difference between the calculated energies with and without considering ZPE was discussed in Supplemental Fig. S1. 
 In addition to the known AE-metal borides \cite{esfahani2017novel,shah2013stability,kolmogorov2012pressure,cui2022superconducting}, we uncovered several new thermodynamically stable M$_{x}$B$_{y}$ compounds, including Mg$_{3}$B$_{8}$, Ca$_{2}$B$_{5}$, Ca$_{3}$B$_{5}$, Sr$_{2}$B$_{5}$, and BaB with $C$2/$m$, $P$6$_{3}$/$mmc$, $P$6$_{3}$/$mmc$, $F$$\overline{4}$3$m$, and $R$$\overline{3}$$m$ structure, respectively, where Sr$_{2}$B$_{5}$ further transforms into $P$6$_{3}$/$mmc$ phase at 44 GPa (The detailed structural information is listed in Table S1 \cite{SI}). The absence of imaginary frequency in the phonon spectra indicates that these borides are also dynamically stable (Fig. S2 \cite{SI}), confirming the robustness of the stability. By examining the structures, we found that, except for $F$$\overline{4}$3$m$-Sr$_{2}$B$_{5}$, all the M$_{x}$B$_{y}$ compounds adopt layered-like morphology with alternatively stacked boron and metal atomic layers (Fig. S3 \cite{SI}). The boron atoms should be mainly $sp^{2}$-hybridized in these M$_{x}$B$_{y}$ compounds like other layered metal borides \cite{wu2016coexistence,cui2022superconducting,shah2013stability}. On the other hand, $F$$\overline{4}$3$m$-Sr$_{2}$B$_{5}$ exhibits a unique three-dimensional boron framework as presented in Fig. \ref{fig:str}. It consists of two nonequivalent boron atoms, namely B$_{D}$ and  B$_{T}$, occupying 4\textit{c} (0.25, 0.25, 0.25) and 16\textit{e} (0.59, 0.59, 0.41) Wyckoff position, respectively, where both of them are four coordinated. Specifically, B$_{D}$ is surrounded by four B$_{T}$ atoms, which forms a regular tetrahedron B$_{5}$ unit similar to C atoms in diamond, indicating B$_{D}$ is $sp^{3}$-hybridized. B$_{T}$ has three shorter B$_{T}$-B$_{T}$ bonds and one slightly longer B$_{T}$-B$_{D}$ bond and forms an irregular tetrahedron B$_{5}^{'}$ unit, in which B$_{T}$ atoms also form a regular tetrahedron B$_{4}$ unit by itself. This is close to the motif of C atoms in T-carbon \cite{sheng2011t,zhang2017pseudo}, implying B$_{T}$ exhibits resembling anisotropic $sp^{3}$-hybridization due to the unevenly distributed bond lengths. Further electron localization function (ELF) and crystal orbital Hamiltonian population (COHP) calculations confirm the strong covalent nature between B atoms (Fig. S4 \cite{SI}). Meanwhile, Sr-B is ionically bonded since electrons are transferred from Sr to B. Remarkably, the boron sublattice of $F$$\overline{4}$3$m$-Sr$_{2}$B$_{5}$ can be viewed as isostructure to diamond, where one of the nonequivalent atomic position is occupied by B$_{4}$ unit instead of B atom. Thus, to the best of our knowledge, $F$$\overline{4}$3$m$-Sr$_{2}$B$_{5}$ should be the first metal boride exhibits fully $sp^{3}$-hybridized $\sigma$-bonded boron framework that might be beneficial to the superconductivity.

\begin{figure}[!t]
	\begin{center}
		\epsfxsize=8.5cm

  \epsffile{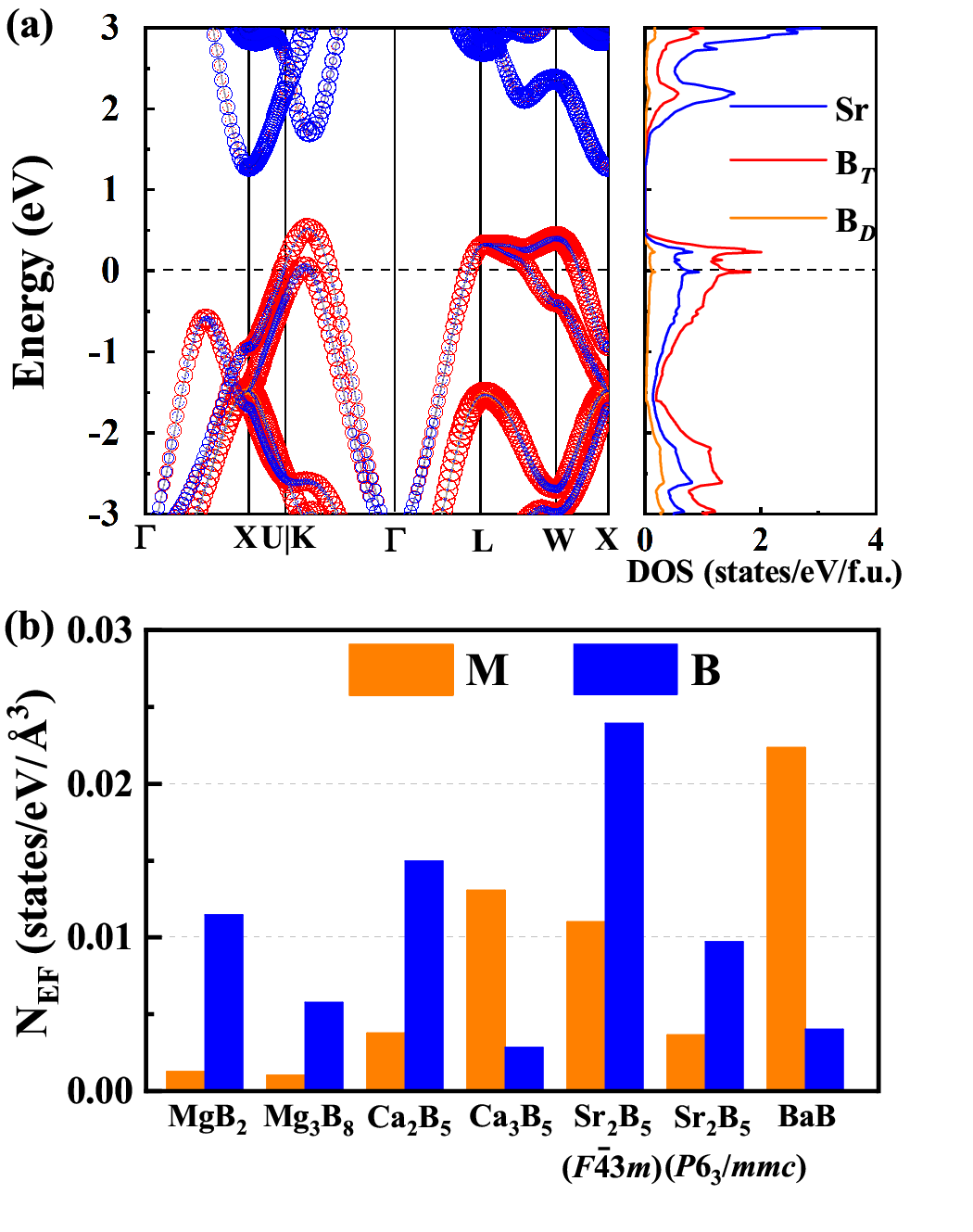}
	\end{center}
	\caption{(Color online) (a) Band structure and DOS of $F$$\overline{4}$3$m$ Sr$_{2}$B$_{5}$ at 40 GPa.  (b) The projected density of states (PDOS) of boron atoms at the Fermi level in MgB$_{2}$, Mg$_{3}$B$_{8}$, Ca$_{2}$B$_{5}$, Ca$_{3}$B$_{5}$, $F$$\overline{4}$3$m$  Sr$_{2}$B$_{5}$, $P$6$_{3}$/$mmc$ Sr$_{2}$B$_{5}$, and BaB at 0, 50 ,30, 40, 40, 50, and 80 GPa, respectively. }
	\label{fig:ele}
\end{figure}
We next examined the electronic properties of the AE-metal borides by modeling the band structures and density of states (DOS) as shown in Fig. 3 and Fig. S5 \cite{SI}. Clearly,  the newly predicted metal borides are all metallic. Based on the boron hybridization scheme, these borides exhibit different band structure characteristics. The layered-like M$_{x}$B$_{y}$ compounds have both  $\sigma$ and $\pi$ bands crossing the Fermi level and $F$$\overline{4}$3$m$-Sr$_{2}$B$_{5}$ only has $\sigma$ conducting bands, analogous to that of MgB$_{2}$ and hole-doped diamond, respectively. Besides the metal-rich Ca$_{3}$B$_{5}$ and BaB, boron dominates the contribution to the DOS at the Fermi level (N$_{Ef}$) due to the charge transfer from metal to boron atoms, as deciphered in Fig. \ref{fig:ele}b. In the case of metal-rich borides, the amount of electrons provided by the metal atoms exceeds the needs of fully shelled boron atoms. Thus,  N$_{Ef}$ is dominated by metal atoms. Importantly, among these metal borides, $F$$\overline{4}$3$m$ Sr$_{2}$B$_{5}$ exhibits not only the highest total N$_{Ef}$ but also the highest boron contribution, which is almost twice as much as that of MgB$_{2}$. This is because the Fermi level is located right at the peak position of the DOS, corresponding to a van Hove singularity along the K-$\Gamma$ direction. The projected DOS further reveals that the electrons in the vicinity of the Fermi level are coming from B$_{T}$, where B$_{D}$  merely contributes to the N$_{Ef}$ because it is nearly fully shelled due to the ideal $sp^{3}$-hybridization. The high N$_{Ef}$ together with the solely $\sigma$ characteristic of boron suggest $F$$\overline{4}$3$m$ Sr$_{2}$B$_{5}$ should be a potential high-temperature superconductor if $\sigma$ electronic bands  couple strongly to the corresponding phonon modes. We also noticed that Ca$_{2}$B$_{5}$ has a boron contributed N$_{Ef}$ higher than that of MgB$_{2}$. However, it is a mixture of $\sigma$ and $\pi$ electrons. The superconductivity of Ca$_{2}$B$_{5}$ might not be able to surpass that of MgB$_{2}$.

\begin{figure}[!t]
	\begin{center}
		\epsfxsize=8.5cm
		\epsffile{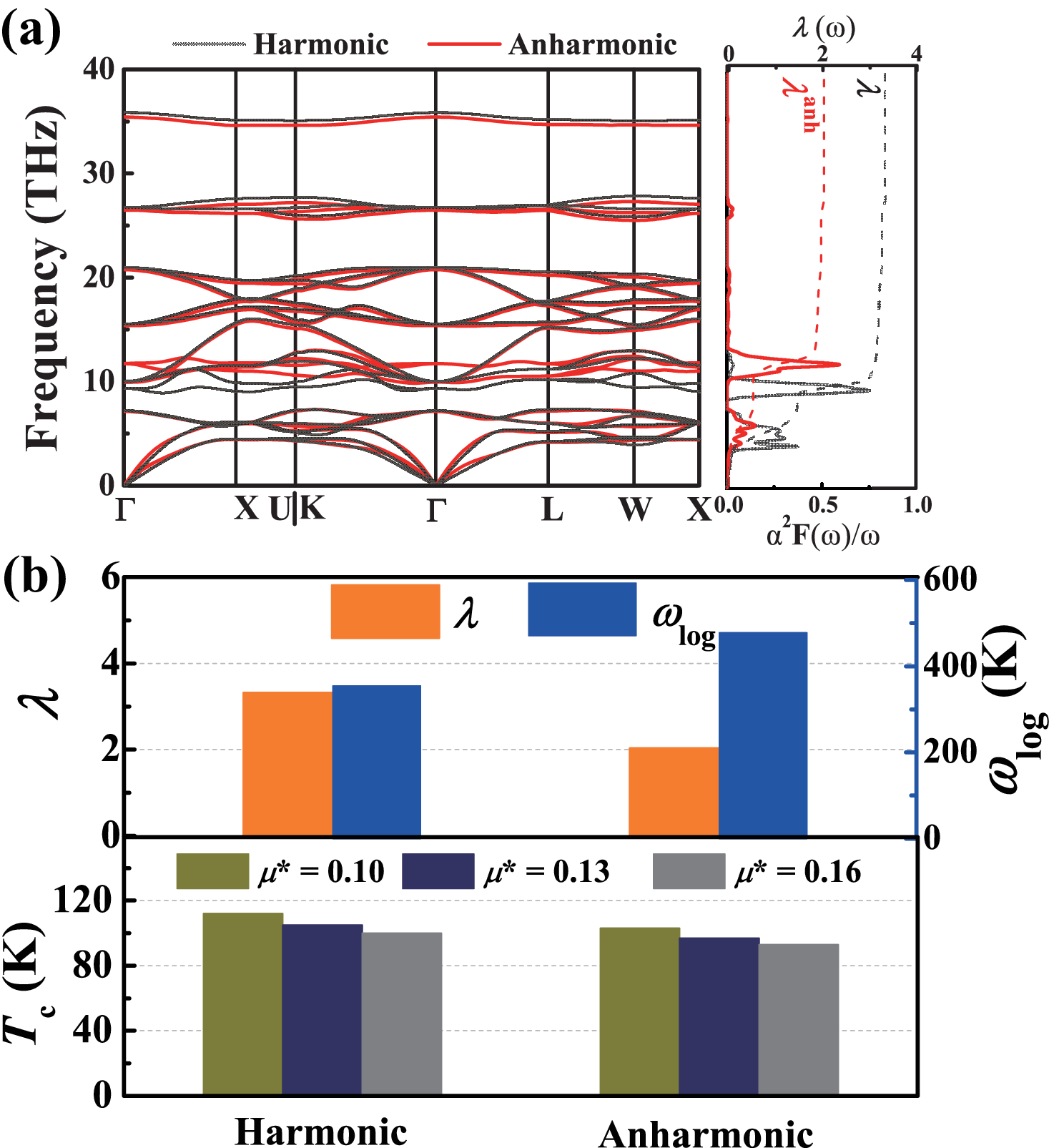}
	\end{center}
	\caption{(Color online) (a) Phonon dispersion relations, Eliashberg spectral function, (b) the electron-phonon coupling parameter $\lambda$, the logarithmic average frequency $\omega_\mathrm{log}$ and the superconducting temperature $T_\mathrm{c}$ of $F$$\overline{4}$3$m$  Sr$_{2}$B$_{5}$ by performing harmonic and anharmonic calculations at  40 GPa. }
	\label{fig:phon}
\end{figure}

To investigate the potential superconductivity in these AE-metal borides, electron-phonon coupling simulations were performed. Encouragingly, the estimated EPC parameters, $\lambda$, of $F$$\overline{4}$3$m$ Sr$_{2}$B$_{5}$ is as high as 3.32 at 40 GPa. This value not only is nearly 3-4 times higher than that of MgB$_{2}$ but also higher than that of any $sp^{3}$-hybridized borocarbides studied so far and close to that of high-$T_\mathrm{c}$ clathrate superhyrides. As depicted in Fig. 4(a), the large coupling strength is derived primarily from the $T_{2}$ and \textit{E} modes over the whole Brillouin zone, which corresponded to the stretching of Sr-B bonds (2-7 THz) and rocking vibrations of B$_4$ units (8-10 THz). The integral ($\lambda$($\omega$)) of the frequency divided Eliashberg phonon spectral function ($\alpha^2F(\omega)/\omega$) suggests these two modes contribute almost equally to the overall $\lambda$. This phenomenon is different from MgB$_{2}$ and metal borocarbides but similar to $\alpha$-MoB$_{2}$ as metal atom shows pronounced contribution to N$_{Ef}$. Since the electron-phonon coupling is quite strong in $F$$\overline{4}$3$m$ Sr$_{2}$B$_{5}$ ($\lambda$ > 1.5), the $T_\mathrm{c}$ values and superconducting energy gaps ($\Delta$) were evaluated through the direct numerical solutions of the anisotropic Migdal-Eliashberg equations \cite{eliashberg1960interactions,sanna2018ab,giustino2007electron} using the calculated logarithmic average frequency ($\omega_\mathrm{log}$) and typical Coulomb potential parameters ($\mu$*) of 0.1, 0.13, and 0.16 . The estimated single $\sigma$ gap is about 22 meV (Fig. S8 \cite{SI}) and $T_\mathrm{c}$ value is 105 K using $\mu$* = 0.13 for $F$$\overline{4}$3$m$ Sr$_{2}$B$_{5}$ at 40 GPa (Fig. \ref{fig:phon}), tripling that of MgB$_{2}$ and validating the observation of $\lambda$ (those for $\mu$* = 0.1, 0.16 are also provided in Fig.4 and Table S3 \cite{SI}). This predicted $T_\mathrm{c}$ certainly exceeds the boiling point of liquid-nitrogen (77 K), setting a new superconducting record among all known thermodynamically stable metal borides to the best of our knowledge.

Given the low mass of B, previous studies have demonstrated that anharmonicity played an important role in determining the superconductivity of MgB$_{2}$, especially for the $E_{2g}$ phonons \cite{liu2001beyond,yildirim2001giant,lazzeri2003anharmonic}. We then carried out additional EPC calculations to investigate the occurrence of anharmonic effects using the stochastic self-consistent harmonic approximation \cite{errea2014anharmonic}. As shown in Fig. \ref{fig:phon}, the anharmonic correction only hardens and broadens the $B_4$ rocking \textit{E} modes, while other phonon modes barely change, similar to that of MgB$_{2}$. The average phonon frequency of \textit{E} modes increases from 9.3 THz to 11.6 THz and the width of \textit{E} modes is increased by 1 THz. Meanwhile, \textit{E} modes become the major contributor to the anharmonic Eliashberg function, accounting for 71$\%$ of total $\lambda^\mathrm{anh}$, while the contribution of $T_{2}$ modes reduces to 25$\%$, indicating the superconductivity of $F$$\overline{4}$3$m$ Sr$_{2}$B$_{5}$ should be mainly determined by the coupling between $\sigma$ bands and \textit{E} phonons. This results in a reduction of the anharmonic EPC parameter $\lambda^\mathrm{anh}$ to 2.02, 40$\%$ smaller than the harmonic one. In contrast, the anharmonicity enhances the logarithmic average frequency $\omega_\mathrm{log}$ by 34$\%$ from 354 K to 477 K. Consequently, the anharmonic $T_\mathrm{c}^\mathrm{anh}$ value is dropped by 10$\%$ to 95 K, which still is quite high.

As for other newly predicted borides, only Ca$_2$B$_5$, $P$6$_{3}$/$mmc$ Sr$_2$B$_5$, and BaB exhibit superconductivity, with calculated $T_\mathrm{c}$ values of 3.6 K, 6.4 K, and 3.9 K at 30, 50, and 80 GPa, respectively. The details of superconducting properties are shown in Table S4 \cite{SI}.

The high-temperature superconductivity of $F$$\overline{4}$3$m$ Sr$_{2}$B$_{5}$ drives us to further evaluate its synthesizability. As shown in Fig. S10 \cite{SI}, the pressure-temperature phase diagram of Sr$_{2}$B$_{5}$ was constructed up to 100 GPa and 2000 K via harmonic free-energy calculations since most of the metal borides are synthesized at high temperatures \cite{friedrich2011synthesis}. Clearly, once temperature effects are considered, the stable range of $F$$\overline{4}$3$m$ Sr$_{2}$B$_{5}$ increases from 38-44 GPa at 0 K to 34-65 GPa at 2000 K, suggesting high temperature is good for the stabilization of $F$$\overline{4}$3$m$ Sr$_{2}$B$_{5}$. Thus, $F$$\overline{4}$3$m$ Sr$_{2}$B$_{5}$ could be synthesized at moderate pressure by the large-volume press (LVP) since its stable range lies within the operating pressure and temperature range of LVP \cite{ishii2019breakthrough}. Compared to high-$T_\mathrm{c}$ BCS hydride superconductors, which can only be realized by the diamond anvil cell with micrometer-sized samples, the production of $F$$\overline{4}$3$m$ Sr$_{2}$B$_{5}$ shows a certain advantage as LVP can provide much larger millimeter-sized samples for structure characterization and properties measurements. Moreover, we also found $F$$\overline{4}$3$m$ Sr$_{2}$B$_{5}$ is dynamically stable down to ambient pressure with anharmonic corrections (Fig. S11 \cite{SI}), implying it might be able to retain to 0 GPa with the proper quenching method. The $T_\mathrm{c}^\mathrm{anh}$ of $F$$\overline{4}$3$m$ Sr$_{2}$B$_{5}$ at 0 GPa is predicted to be even higher around 115 K (Fig. S12 \cite{SI}).

As part of efforts to develop high-temperature BCS superconductors, we designed a promising thermodynamically stable $F$$\overline{4}$3$m$ Sr$_{2}$B$_{5}$ with record-high $T_\mathrm{c}$ at moderate synthetic pressure ($\sim$ 40 GPa). We note this is the first fully metallic $sp^{3}$-hybridized boron framework example of metal borides, and we emphasize that the established  B-B $\sigma$ bonding is a key to achieving high-$T_\mathrm{c}$ at low pressures. For instance, if B$_{D}$ atom is also replaced by the strongly coupled $B_4$ unit in $F$$\overline{4}$3$m$ Sr$_{2}$B$_{5}$, the $T_\mathrm{c}$ might be even higher. Further investigation of other metal borides and lightweight compounds consisting of similar $\sigma$-bonding frameworks is expected which may lead to the discovery of potential high-$T_\mathrm{c}$ BCS superconductor (> 77 K) at ambient pressure.

In conclusion, our extensive first-principles structure searches for AE-metal (Mg, Ca, Sr, Ba) borides have revealed the appearance of stable Mg$_{3}$B$_{8}$, Ca$_{2}$B$_{5}$, Ca$_{3}$B$_{5}$, Sr$_{2}$B$_{5}$, and BaB below 100 GPa. Among them, $F$$\overline{4}$3$m$ Sr$_{2}$B$_{5}$ consisting of fully metallic $sp^{3}$-hybridized B-B $\sigma$ bonds exhibits potential high-$T_\mathrm{c}$ of 105 K at 40 GPa, with the possibility retaining to ambient pressure, that originates from the strong coupling between $\sigma$ electronic bands and rocking vibrations of $B_4$ units. This unique boron framework resembles the diamond structure, serving as a design guidance for future investigation. Our results provide a useful structural prototype for designing high-temperature superconductors at low pressures and will stimulate further experimental synthesis and theoretical predictions for a large variety of metallic $sp^{3}$-hybridized lightweight compounds.

This work is supported by by National Key Research and Development Program of China (Grant No. 2022YFA1402304), the National Natural Science Foundation of China (Grant no. 12022408, 12374008, 1202290458, 52288102, 52090024, 12074138, 12374007), the Interdisciplinary Integration and Innovation Project of JLU, Fundamental Research Funds for the Central Universities, Program for Jilin University Science and Technology Innovative Research Team (2021TD–05), Jilin Province Science and Technology Development Program (Grant No. YDZJ202102CXJD016), the Strategic Priority Research Program of Chinese Academy of Sciences (Grant No. XDB33000000) and computing facilities at the High-Performance Computing Centre of Jilin University.


\begin{thebibliography}{46}
	\expandafter\ifx\csname natexlab\endcsname\relax\def\natexlab#1{#1}\fi
	\expandafter\ifx\csname bibnamefont\endcsname\relax
	\def\bibnamefont#1{#1}\fi
	\expandafter\ifx\csname bibfnamefont\endcsname\relax
	\def\bibfnamefont#1{#1}\fi
	\expandafter\ifx\csname citenamefont\endcsname\relax
	\def\citenamefont#1{#1}\fi
	\expandafter\ifx\csname url\endcsname\relax
	\def\url#1{\texttt{#1}}\fi
	\expandafter\ifx\csname urlprefix\endcsname\relax\def\urlprefix{URL }\fi
	\providecommand{\bibinfo}[2]{#2}
	\providecommand{\eprint}[2][]{\url{#2}}
	
	\bibitem[{\citenamefont{Onnes}(1911)}]{1911}
	\bibinfo{author}{\bibfnamefont{H.~K.} \bibnamefont{Onnes}},
	\bibinfo{journal}{Commun. Phys. Lab. Univ. Leiden, b}
	\textbf{\bibinfo{volume}{120}} (\bibinfo{year}{1911}).
	
	\bibitem[{\citenamefont{Suhl et~al.}(1959)\citenamefont{Suhl, Matthias, and
			Walker}}]{suhl1959bardeen}
	\bibinfo{author}{\bibfnamefont{H.}~\bibnamefont{Suhl}},
	\bibinfo{author}{\bibfnamefont{B.}~\bibnamefont{Matthias}}, \bibnamefont{and}
	\bibinfo{author}{\bibfnamefont{L.}~\bibnamefont{Walker}},
	\bibinfo{journal}{Phys. Rev. Lett.} \textbf{\bibinfo{volume}{3}},
	\bibinfo{pages}{552} (\bibinfo{year}{1959}).
	
	\bibitem[{\citenamefont{Drozdov et~al.}(2015)\citenamefont{Drozdov, Eremets,
			Troyan, Ksenofontov, and Shylin}}]{drozdov2015conventional}
	\bibinfo{author}{\bibfnamefont{A.}~\bibnamefont{Drozdov}},
	\bibinfo{author}{\bibfnamefont{M.}~\bibnamefont{Eremets}},
	\bibinfo{author}{\bibfnamefont{I.}~\bibnamefont{Troyan}},
	\bibinfo{author}{\bibfnamefont{V.}~\bibnamefont{Ksenofontov}},
	\bibnamefont{and} \bibinfo{author}{\bibfnamefont{S.~I.}
		\bibnamefont{Shylin}}, \bibinfo{journal}{Nature}
	\textbf{\bibinfo{volume}{525}}, \bibinfo{pages}{73} (\bibinfo{year}{2015}).
	
	\bibitem[{\citenamefont{Peng et~al.}(2017)\citenamefont{Peng, Sun, Pickard,
			Needs, Wu, and Ma}}]{peng2017hydrogen}
	\bibinfo{author}{\bibfnamefont{F.}~\bibnamefont{Peng}},
	\bibinfo{author}{\bibfnamefont{Y.}~\bibnamefont{Sun}},
	\bibinfo{author}{\bibfnamefont{C.~J.} \bibnamefont{Pickard}},
	\bibinfo{author}{\bibfnamefont{R.~J.} \bibnamefont{Needs}},
	\bibinfo{author}{\bibfnamefont{Q.}~\bibnamefont{Wu}}, \bibnamefont{and}
	\bibinfo{author}{\bibfnamefont{Y.}~\bibnamefont{Ma}}, \bibinfo{journal}{Phys.
		Rev. Lett.} \textbf{\bibinfo{volume}{119}}, \bibinfo{pages}{107001}
	(\bibinfo{year}{2017}).
	
	\bibitem[{\citenamefont{Drozdov et~al.}(2019)\citenamefont{Drozdov, Kong,
			Minkov, Besedin, Kuzovnikov, Mozaffari, Balicas, Balakirev, Graf, Prakapenka
			et~al.}}]{drozdov2019superconductivity}
	\bibinfo{author}{\bibfnamefont{A.}~\bibnamefont{Drozdov}},
	\bibinfo{author}{\bibfnamefont{P.}~\bibnamefont{Kong}},
	\bibinfo{author}{\bibfnamefont{V.}~\bibnamefont{Minkov}},
	\bibinfo{author}{\bibfnamefont{S.}~\bibnamefont{Besedin}},
	\bibinfo{author}{\bibfnamefont{M.}~\bibnamefont{Kuzovnikov}},
	\bibinfo{author}{\bibfnamefont{S.}~\bibnamefont{Mozaffari}},
	\bibinfo{author}{\bibfnamefont{L.}~\bibnamefont{Balicas}},
	\bibinfo{author}{\bibfnamefont{F.}~\bibnamefont{Balakirev}},
	\bibinfo{author}{\bibfnamefont{D.}~\bibnamefont{Graf}},
	\bibinfo{author}{\bibfnamefont{V.}~\bibnamefont{Prakapenka}},
	\bibnamefont{et~al.}, \bibinfo{journal}{Nature}
	\textbf{\bibinfo{volume}{569}}, \bibinfo{pages}{528} (\bibinfo{year}{2019}).
	
	\bibitem[{\citenamefont{Somayazulu et~al.}(2019)\citenamefont{Somayazulu,
			Ahart, Mishra, Geballe, Baldini, Meng, Struzhkin, and
			Hemley}}]{somayazulu2019evidence}
	\bibinfo{author}{\bibfnamefont{M.}~\bibnamefont{Somayazulu}},
	\bibinfo{author}{\bibfnamefont{M.}~\bibnamefont{Ahart}},
	\bibinfo{author}{\bibfnamefont{A.~K.} \bibnamefont{Mishra}},
	\bibinfo{author}{\bibfnamefont{Z.~M.} \bibnamefont{Geballe}},
	\bibinfo{author}{\bibfnamefont{M.}~\bibnamefont{Baldini}},
	\bibinfo{author}{\bibfnamefont{Y.}~\bibnamefont{Meng}},
	\bibinfo{author}{\bibfnamefont{V.~V.} \bibnamefont{Struzhkin}},
	\bibnamefont{and} \bibinfo{author}{\bibfnamefont{R.~J.}
		\bibnamefont{Hemley}}, \bibinfo{journal}{Phys. Rev. Lett.}
	\textbf{\bibinfo{volume}{122}}, \bibinfo{pages}{027001}
	(\bibinfo{year}{2019}).
	
	\bibitem[{\citenamefont{Ma et~al.}(2022)\citenamefont{Ma, Wang, Xie, Yang,
			Wang, Zhou, Liu, Yu, Zhao, Wang et~al.}}]{ma2022high}
	\bibinfo{author}{\bibfnamefont{L.}~\bibnamefont{Ma}},
	\bibinfo{author}{\bibfnamefont{K.}~\bibnamefont{Wang}},
	\bibinfo{author}{\bibfnamefont{Y.}~\bibnamefont{Xie}},
	\bibinfo{author}{\bibfnamefont{X.}~\bibnamefont{Yang}},
	\bibinfo{author}{\bibfnamefont{Y.}~\bibnamefont{Wang}},
	\bibinfo{author}{\bibfnamefont{M.}~\bibnamefont{Zhou}},
	\bibinfo{author}{\bibfnamefont{H.}~\bibnamefont{Liu}},
	\bibinfo{author}{\bibfnamefont{X.}~\bibnamefont{Yu}},
	\bibinfo{author}{\bibfnamefont{Y.}~\bibnamefont{Zhao}},
	\bibinfo{author}{\bibfnamefont{H.}~\bibnamefont{Wang}}, \bibnamefont{et~al.},
	\bibinfo{journal}{Phys. Rev. Lett.} \textbf{\bibinfo{volume}{128}},
	\bibinfo{pages}{167001} (\bibinfo{year}{2022}).
	
	\bibitem[{\citenamefont{Li et~al.}(2014)\citenamefont{Li, Hao, Liu, Li, and
			Ma}}]{li2014metallization}
	\bibinfo{author}{\bibfnamefont{Y.}~\bibnamefont{Li}},
	\bibinfo{author}{\bibfnamefont{J.}~\bibnamefont{Hao}},
	\bibinfo{author}{\bibfnamefont{H.}~\bibnamefont{Liu}},
	\bibinfo{author}{\bibfnamefont{Y.}~\bibnamefont{Li}}, \bibnamefont{and}
	\bibinfo{author}{\bibfnamefont{Y.}~\bibnamefont{Ma}}, \bibinfo{journal}{J.
		Chem. Phys.} \textbf{\bibinfo{volume}{140}}, \bibinfo{pages}{174712}
	(\bibinfo{year}{2014}).
	
	\bibitem[{\citenamefont{Liu et~al.}(2017)\citenamefont{Liu, Naumov, Hoffmann,
			Ashcroft, and Hemley}}]{liu2017potential}
	\bibinfo{author}{\bibfnamefont{H.}~\bibnamefont{Liu}},
	\bibinfo{author}{\bibfnamefont{I.~I.} \bibnamefont{Naumov}},
	\bibinfo{author}{\bibfnamefont{R.}~\bibnamefont{Hoffmann}},
	\bibinfo{author}{\bibfnamefont{N.}~\bibnamefont{Ashcroft}}, \bibnamefont{and}
	\bibinfo{author}{\bibfnamefont{R.~J.} \bibnamefont{Hemley}},
	\bibinfo{journal}{Proc. Natl. Acad. Sci. U.S.A.}
	\textbf{\bibinfo{volume}{114}}, \bibinfo{pages}{6990} (\bibinfo{year}{2017}).
	
	\bibitem[{\citenamefont{Wang et~al.}(2012{\natexlab{a}})\citenamefont{Wang,
			Tse, Tanaka, Iitaka, and Ma}}]{wang2012superconductive}
	\bibinfo{author}{\bibfnamefont{H.}~\bibnamefont{Wang}},
	\bibinfo{author}{\bibfnamefont{J.~S.} \bibnamefont{Tse}},
	\bibinfo{author}{\bibfnamefont{K.}~\bibnamefont{Tanaka}},
	\bibinfo{author}{\bibfnamefont{T.}~\bibnamefont{Iitaka}}, \bibnamefont{and}
	\bibinfo{author}{\bibfnamefont{Y.}~\bibnamefont{Ma}}, \bibinfo{journal}{Proc.
		Natl. Acad. Sci. U.S.A.} \textbf{\bibinfo{volume}{109}},
	\bibinfo{pages}{6463} (\bibinfo{year}{2012}{\natexlab{a}}).
	
	\bibitem[{\citenamefont{Nagamatsu et~al.}(2001)\citenamefont{Nagamatsu,
			Nakagawa, Muranaka, Zenitani, and Akimitsu}}]{nagamatsu2001superconductivity}
	\bibinfo{author}{\bibfnamefont{J.}~\bibnamefont{Nagamatsu}},
	\bibinfo{author}{\bibfnamefont{N.}~\bibnamefont{Nakagawa}},
	\bibinfo{author}{\bibfnamefont{T.}~\bibnamefont{Muranaka}},
	\bibinfo{author}{\bibfnamefont{Y.}~\bibnamefont{Zenitani}}, \bibnamefont{and}
	\bibinfo{author}{\bibfnamefont{J.}~\bibnamefont{Akimitsu}},
	\bibinfo{journal}{Nature} \textbf{\bibinfo{volume}{410}}, \bibinfo{pages}{63}
	(\bibinfo{year}{2001}).
	
	\bibitem[{\citenamefont{Kortus et~al.}(2001)\citenamefont{Kortus, Mazin,
			Belashchenko, Antropov, and Boyer}}]{kortus2001superconductivity}
	\bibinfo{author}{\bibfnamefont{J.}~\bibnamefont{Kortus}},
	\bibinfo{author}{\bibfnamefont{I.}~\bibnamefont{Mazin}},
	\bibinfo{author}{\bibfnamefont{K.~D.} \bibnamefont{Belashchenko}},
	\bibinfo{author}{\bibfnamefont{V.~P.} \bibnamefont{Antropov}},
	\bibnamefont{and} \bibinfo{author}{\bibfnamefont{L.}~\bibnamefont{Boyer}},
	\bibinfo{journal}{Phys. Rev. Lett.} \textbf{\bibinfo{volume}{86}},
	\bibinfo{pages}{4656} (\bibinfo{year}{2001}).
	
	\bibitem[{\citenamefont{Fisk et~al.}(1971)\citenamefont{Fisk, Lawsom, Matthias,
			and Corenzwit}}]{fisk1971superconducting}
	\bibinfo{author}{\bibfnamefont{Z.}~\bibnamefont{Fisk}},
	\bibinfo{author}{\bibfnamefont{A.}~\bibnamefont{Lawsom}},
	\bibinfo{author}{\bibfnamefont{B.}~\bibnamefont{Matthias}}, \bibnamefont{and}
	\bibinfo{author}{\bibfnamefont{E.}~\bibnamefont{Corenzwit}},
	\bibinfo{journal}{Phys. Lett. A} \textbf{\bibinfo{volume}{37}},
	\bibinfo{pages}{251} (\bibinfo{year}{1971}).
	
	\bibitem[{\citenamefont{Schneider et~al.}(1987)\citenamefont{Schneider, Geerk,
			and Rietschel}}]{schneider1987electron}
	\bibinfo{author}{\bibfnamefont{R.}~\bibnamefont{Schneider}},
	\bibinfo{author}{\bibfnamefont{J.}~\bibnamefont{Geerk}}, \bibnamefont{and}
	\bibinfo{author}{\bibfnamefont{H.}~\bibnamefont{Rietschel}},
	\bibinfo{journal}{Europhys. Lett.} \textbf{\bibinfo{volume}{4}},
	\bibinfo{pages}{845} (\bibinfo{year}{1987}).
	
	\bibitem[{\citenamefont{Lortz et~al.}(2006)\citenamefont{Lortz, Wang, Tutsch,
			Abe, Meingast, Popovich, Knafo, Shitsevalova, Paderno, and
			Junod}}]{lortz2006superconductivity}
	\bibinfo{author}{\bibfnamefont{R.}~\bibnamefont{Lortz}},
	\bibinfo{author}{\bibfnamefont{Y.}~\bibnamefont{Wang}},
	\bibinfo{author}{\bibfnamefont{U.}~\bibnamefont{Tutsch}},
	\bibinfo{author}{\bibfnamefont{S.}~\bibnamefont{Abe}},
	\bibinfo{author}{\bibfnamefont{C.}~\bibnamefont{Meingast}},
	\bibinfo{author}{\bibfnamefont{P.}~\bibnamefont{Popovich}},
	\bibinfo{author}{\bibfnamefont{W.}~\bibnamefont{Knafo}},
	\bibinfo{author}{\bibfnamefont{N.}~\bibnamefont{Shitsevalova}},
	\bibinfo{author}{\bibfnamefont{Y.~B.} \bibnamefont{Paderno}},
	\bibnamefont{and} \bibinfo{author}{\bibfnamefont{A.}~\bibnamefont{Junod}},
	\bibinfo{journal}{Phys. Rev. B} \textbf{\bibinfo{volume}{73}},
	\bibinfo{pages}{024512} (\bibinfo{year}{2006}).
	
	\bibitem[{\citenamefont{Akopov et~al.}(2017)\citenamefont{Akopov, Yeung, and
			Kaner}}]{akopov2017rediscovering}
	\bibinfo{author}{\bibfnamefont{G.}~\bibnamefont{Akopov}},
	\bibinfo{author}{\bibfnamefont{M.~T.} \bibnamefont{Yeung}}, \bibnamefont{and}
	\bibinfo{author}{\bibfnamefont{R.~B.} \bibnamefont{Kaner}},
	\bibinfo{journal}{Adv. Mater.} \textbf{\bibinfo{volume}{29}},
	\bibinfo{pages}{1604506} (\bibinfo{year}{2017}).
	
	\bibitem[{\citenamefont{Pei et~al.}(2023)\citenamefont{Pei, Zhang, Wang, Zhao,
			Gao, Gong, Tian, Luo, Li, Yang et~al.}}]{pei2023pressure}
	\bibinfo{author}{\bibfnamefont{C.}~\bibnamefont{Pei}},
	\bibinfo{author}{\bibfnamefont{J.}~\bibnamefont{Zhang}},
	\bibinfo{author}{\bibfnamefont{Q.}~\bibnamefont{Wang}},
	\bibinfo{author}{\bibfnamefont{Y.}~\bibnamefont{Zhao}},
	\bibinfo{author}{\bibfnamefont{L.}~\bibnamefont{Gao}},
	\bibinfo{author}{\bibfnamefont{C.}~\bibnamefont{Gong}},
	\bibinfo{author}{\bibfnamefont{S.}~\bibnamefont{Tian}},
	\bibinfo{author}{\bibfnamefont{R.}~\bibnamefont{Luo}},
	\bibinfo{author}{\bibfnamefont{M.}~\bibnamefont{Li}},
	\bibinfo{author}{\bibfnamefont{W.}~\bibnamefont{Yang}}, \bibnamefont{et~al.},
	\bibinfo{journal}{Natl. Sci. Rev.} \textbf{\bibinfo{volume}{10}},
	\bibinfo{pages}{nwad034} (\bibinfo{year}{2023}).
	
	\bibitem[{\citenamefont{Wu et~al.}(2016)\citenamefont{Wu, Wan, Liu, Gou, Yao,
			Li, Zhang, Gao, and Mao}}]{wu2016coexistence}
	\bibinfo{author}{\bibfnamefont{L.}~\bibnamefont{Wu}},
	\bibinfo{author}{\bibfnamefont{B.}~\bibnamefont{Wan}},
	\bibinfo{author}{\bibfnamefont{H.}~\bibnamefont{Liu}},
	\bibinfo{author}{\bibfnamefont{H.}~\bibnamefont{Gou}},
	\bibinfo{author}{\bibfnamefont{Y.}~\bibnamefont{Yao}},
	\bibinfo{author}{\bibfnamefont{Z.}~\bibnamefont{Li}},
	\bibinfo{author}{\bibfnamefont{J.}~\bibnamefont{Zhang}},
	\bibinfo{author}{\bibfnamefont{F.}~\bibnamefont{Gao}}, \bibnamefont{and}
	\bibinfo{author}{\bibfnamefont{H.-K.} \bibnamefont{Mao}},
	\bibinfo{journal}{J. Phys. Chem. Lett.} \textbf{\bibinfo{volume}{7}},
	\bibinfo{pages}{4898} (\bibinfo{year}{2016}).
	
	\bibitem[{\citenamefont{Cui et~al.}(2022)\citenamefont{Cui, Yang, Qu, Zhang,
			Liu, and Yang}}]{cui2022superconducting}
	\bibinfo{author}{\bibfnamefont{Z.}~\bibnamefont{Cui}},
	\bibinfo{author}{\bibfnamefont{Q.}~\bibnamefont{Yang}},
	\bibinfo{author}{\bibfnamefont{X.}~\bibnamefont{Qu}},
	\bibinfo{author}{\bibfnamefont{X.}~\bibnamefont{Zhang}},
	\bibinfo{author}{\bibfnamefont{Y.}~\bibnamefont{Liu}}, \bibnamefont{and}
	\bibinfo{author}{\bibfnamefont{G.}~\bibnamefont{Yang}}, \bibinfo{journal}{J.
		Mater. Chem. C} \textbf{\bibinfo{volume}{10}}, \bibinfo{pages}{672}
	(\bibinfo{year}{2022}).
	
	\bibitem[{\citenamefont{Shah and Kolmogorov}(2013)}]{shah2013stability}
	\bibinfo{author}{\bibfnamefont{S.}~\bibnamefont{Shah}} \bibnamefont{and}
	\bibinfo{author}{\bibfnamefont{A.~N.} \bibnamefont{Kolmogorov}},
	\bibinfo{journal}{Phys. Rev. B} \textbf{\bibinfo{volume}{88}},
	\bibinfo{pages}{014107} (\bibinfo{year}{2013}).
	
	\bibitem[{\citenamefont{Zhao et~al.}(2019)\citenamefont{Zhao, Lian, Zeng, Dai,
			Meng, and Ni}}]{zhao2019two}
	\bibinfo{author}{\bibfnamefont{Y.}~\bibnamefont{Zhao}},
	\bibinfo{author}{\bibfnamefont{C.}~\bibnamefont{Lian}},
	\bibinfo{author}{\bibfnamefont{S.}~\bibnamefont{Zeng}},
	\bibinfo{author}{\bibfnamefont{Z.}~\bibnamefont{Dai}},
	\bibinfo{author}{\bibfnamefont{S.}~\bibnamefont{Meng}}, \bibnamefont{and}
	\bibinfo{author}{\bibfnamefont{J.}~\bibnamefont{Ni}}, \bibinfo{journal}{Phys.
		Rev. B} \textbf{\bibinfo{volume}{100}}, \bibinfo{pages}{094516}
	(\bibinfo{year}{2019}).
	
	\bibitem[{\citenamefont{Zhao et~al.}(2020)\citenamefont{Zhao, Lian, Zeng, Dai,
			Meng, and Ni}}]{zhao2020mgb}
	\bibinfo{author}{\bibfnamefont{Y.}~\bibnamefont{Zhao}},
	\bibinfo{author}{\bibfnamefont{C.}~\bibnamefont{Lian}},
	\bibinfo{author}{\bibfnamefont{S.}~\bibnamefont{Zeng}},
	\bibinfo{author}{\bibfnamefont{Z.}~\bibnamefont{Dai}},
	\bibinfo{author}{\bibfnamefont{S.}~\bibnamefont{Meng}}, \bibnamefont{and}
	\bibinfo{author}{\bibfnamefont{J.}~\bibnamefont{Ni}}, \bibinfo{journal}{Phys.
		Rev. B} \textbf{\bibinfo{volume}{101}}, \bibinfo{pages}{104507}
	(\bibinfo{year}{2020}).
	
	\bibitem[{\citenamefont{Wang et~al.}(2021{\natexlab{a}})\citenamefont{Wang,
			Zeng, Zhao, Wang, and Ni}}]{wang2021three}
	\bibinfo{author}{\bibfnamefont{Z.}~\bibnamefont{Wang}},
	\bibinfo{author}{\bibfnamefont{S.}~\bibnamefont{Zeng}},
	\bibinfo{author}{\bibfnamefont{Y.}~\bibnamefont{Zhao}},
	\bibinfo{author}{\bibfnamefont{X.}~\bibnamefont{Wang}}, \bibnamefont{and}
	\bibinfo{author}{\bibfnamefont{J.}~\bibnamefont{Ni}}, \bibinfo{journal}{Phys.
		Rev. B} \textbf{\bibinfo{volume}{104}}, \bibinfo{pages}{174519}
	(\bibinfo{year}{2021}{\natexlab{a}}).
	
	\bibitem[{\citenamefont{Geng et~al.}(2023)\citenamefont{Geng, Hilleke, Zhu,
			Wang, Strobel, and Zurek}}]{geng2023conventional}
	\bibinfo{author}{\bibfnamefont{N.}~\bibnamefont{Geng}},
	\bibinfo{author}{\bibfnamefont{K.~P.} \bibnamefont{Hilleke}},
	\bibinfo{author}{\bibfnamefont{L.}~\bibnamefont{Zhu}},
	\bibinfo{author}{\bibfnamefont{X.}~\bibnamefont{Wang}},
	\bibinfo{author}{\bibfnamefont{T.~A.} \bibnamefont{Strobel}},
	\bibnamefont{and} \bibinfo{author}{\bibfnamefont{E.}~\bibnamefont{Zurek}},
	\bibinfo{journal}{J. Am. Chem. Soc.}  (\bibinfo{year}{2023}).
	
	\bibitem[{\citenamefont{Gai et~al.}(2022)\citenamefont{Gai, Guo, Yang, Gao,
			Gao, and Lu}}]{gai2022van}
	\bibinfo{author}{\bibfnamefont{T.~T.} \bibnamefont{Gai}},
	\bibinfo{author}{\bibfnamefont{P.~J.} \bibnamefont{Guo}},
	\bibinfo{author}{\bibfnamefont{H.~C.} \bibnamefont{Yang}},
	\bibinfo{author}{\bibfnamefont{Y.}~\bibnamefont{Gao}},
	\bibinfo{author}{\bibfnamefont{M.}~\bibnamefont{Gao}}, \bibnamefont{and}
	\bibinfo{author}{\bibfnamefont{Z.~Y.} \bibnamefont{Lu}},
	\bibinfo{journal}{Phys. Rev. B} \textbf{\bibinfo{volume}{105}},
	\bibinfo{pages}{224514} (\bibinfo{year}{2022}).
	
	\bibitem[{\citenamefont{Ding et~al.}(2022)\citenamefont{Ding, Feng, Jiang,
			Tian, Zhong, Yang, Chen, and Lin}}]{ding2022ambient}
	\bibinfo{author}{\bibfnamefont{H.~B.} \bibnamefont{Ding}},
	\bibinfo{author}{\bibfnamefont{Y.~J.} \bibnamefont{Feng}},
	\bibinfo{author}{\bibfnamefont{M.~J.} \bibnamefont{Jiang}},
	\bibinfo{author}{\bibfnamefont{H.~L.} \bibnamefont{Tian}},
	\bibinfo{author}{\bibfnamefont{G.~H.} \bibnamefont{Zhong}},
	\bibinfo{author}{\bibfnamefont{C.~L.} \bibnamefont{Yang}},
	\bibinfo{author}{\bibfnamefont{X.~J.} \bibnamefont{Chen}}, \bibnamefont{and}
	\bibinfo{author}{\bibfnamefont{H.~Q.} \bibnamefont{Lin}},
	\bibinfo{journal}{Phys. Rev. B} \textbf{\bibinfo{volume}{106}},
	\bibinfo{pages}{104508} (\bibinfo{year}{2022}).
	
	\bibitem[{\citenamefont{Zhu et~al.}(2020)\citenamefont{Zhu, Borstad, Liu,
			Gu{\'n}ka, Guerette, Dolyniuk, Meng, Greenberg, Prakapenka, Chaloux
			et~al.}}]{zhu2020carbon}
	\bibinfo{author}{\bibfnamefont{L.}~\bibnamefont{Zhu}},
	\bibinfo{author}{\bibfnamefont{G.~M.} \bibnamefont{Borstad}},
	\bibinfo{author}{\bibfnamefont{H.}~\bibnamefont{Liu}},
	\bibinfo{author}{\bibfnamefont{P.~A.} \bibnamefont{Gu{\'n}ka}},
	\bibinfo{author}{\bibfnamefont{M.}~\bibnamefont{Guerette}},
	\bibinfo{author}{\bibfnamefont{J.-A.} \bibnamefont{Dolyniuk}},
	\bibinfo{author}{\bibfnamefont{Y.}~\bibnamefont{Meng}},
	\bibinfo{author}{\bibfnamefont{E.}~\bibnamefont{Greenberg}},
	\bibinfo{author}{\bibfnamefont{V.~B.} \bibnamefont{Prakapenka}},
	\bibinfo{author}{\bibfnamefont{B.~L.} \bibnamefont{Chaloux}},
	\bibnamefont{et~al.}, \bibinfo{journal}{Sci. Adv.}
	\textbf{\bibinfo{volume}{6}}, \bibinfo{pages}{eaay8361}
	(\bibinfo{year}{2020}).
	
	\bibitem[{\citenamefont{Zhang et~al.}(2022)\citenamefont{Zhang, Li, Yang, Wang,
			Yao, and Liu}}]{zhang2022path}
	\bibinfo{author}{\bibfnamefont{P.}~\bibnamefont{Zhang}},
	\bibinfo{author}{\bibfnamefont{X.}~\bibnamefont{Li}},
	\bibinfo{author}{\bibfnamefont{X.}~\bibnamefont{Yang}},
	\bibinfo{author}{\bibfnamefont{H.}~\bibnamefont{Wang}},
	\bibinfo{author}{\bibfnamefont{Y.}~\bibnamefont{Yao}}, \bibnamefont{and}
	\bibinfo{author}{\bibfnamefont{H.}~\bibnamefont{Liu}},
	\bibinfo{journal}{Phys. Rev. B} \textbf{\bibinfo{volume}{105}},
	\bibinfo{pages}{094503} (\bibinfo{year}{2022}).
	
	\bibitem[{\citenamefont{Wang et~al.}(2021{\natexlab{b}})\citenamefont{Wang,
			Yan, and Gao}}]{wang2021high}
	\bibinfo{author}{\bibfnamefont{J.-N.} \bibnamefont{Wang}},
	\bibinfo{author}{\bibfnamefont{X.-W.} \bibnamefont{Yan}}, \bibnamefont{and}
	\bibinfo{author}{\bibfnamefont{M.}~\bibnamefont{Gao}},
	\bibinfo{journal}{Phys. Rev. B} \textbf{\bibinfo{volume}{103}},
	\bibinfo{pages}{144515} (\bibinfo{year}{2021}{\natexlab{b}}).
	
	\bibitem[{\citenamefont{Wang et~al.}(2010)\citenamefont{Wang, Lv, Zhu, and
			Ma}}]{wang2010crystal}
	\bibinfo{author}{\bibfnamefont{Y.}~\bibnamefont{Wang}},
	\bibinfo{author}{\bibfnamefont{J.}~\bibnamefont{Lv}},
	\bibinfo{author}{\bibfnamefont{L.}~\bibnamefont{Zhu}}, \bibnamefont{and}
	\bibinfo{author}{\bibfnamefont{Y.}~\bibnamefont{Ma}}, \bibinfo{journal}{Phys.
		Rev. B} \textbf{\bibinfo{volume}{82}}, \bibinfo{pages}{094116}
	(\bibinfo{year}{2010}).
	
	\bibitem[{\citenamefont{Wang et~al.}(2012{\natexlab{b}})\citenamefont{Wang, Lv,
			Zhu, and Ma}}]{wang2012calypso}
	\bibinfo{author}{\bibfnamefont{Y.}~\bibnamefont{Wang}},
	\bibinfo{author}{\bibfnamefont{J.}~\bibnamefont{Lv}},
	\bibinfo{author}{\bibfnamefont{L.}~\bibnamefont{Zhu}}, \bibnamefont{and}
	\bibinfo{author}{\bibfnamefont{Y.}~\bibnamefont{Ma}},
	\bibinfo{journal}{Comput. Phys. Commun.} \textbf{\bibinfo{volume}{183}},
	\bibinfo{pages}{2063} (\bibinfo{year}{2012}{\natexlab{b}}).
	
	\bibitem[{\citenamefont{Esfahani et~al.}(2017)\citenamefont{Esfahani, Zhu,
			Dong, Oganov, Wang, Rakitin, and Zhou}}]{esfahani2017novel}
	\bibinfo{author}{\bibfnamefont{M.~M.~D.} \bibnamefont{Esfahani}},
	\bibinfo{author}{\bibfnamefont{Q.}~\bibnamefont{Zhu}},
	\bibinfo{author}{\bibfnamefont{H.}~\bibnamefont{Dong}},
	\bibinfo{author}{\bibfnamefont{A.~R.} \bibnamefont{Oganov}},
	\bibinfo{author}{\bibfnamefont{S.}~\bibnamefont{Wang}},
	\bibinfo{author}{\bibfnamefont{M.~S.} \bibnamefont{Rakitin}},
	\bibnamefont{and} \bibinfo{author}{\bibfnamefont{X.~F.} \bibnamefont{Zhou}},
	\bibinfo{journal}{Phys. Chem. Chem. Phys.} \textbf{\bibinfo{volume}{19}},
	\bibinfo{pages}{14486} (\bibinfo{year}{2017}).
	
	\bibitem[{\citenamefont{Kolmogorov et~al.}(2012)\citenamefont{Kolmogorov, Shah,
			Margine, Kleppe, and Jephcoat}}]{kolmogorov2012pressure}
	\bibinfo{author}{\bibfnamefont{A.}~\bibnamefont{Kolmogorov}},
	\bibinfo{author}{\bibfnamefont{S.}~\bibnamefont{Shah}},
	\bibinfo{author}{\bibfnamefont{E.}~\bibnamefont{Margine}},
	\bibinfo{author}{\bibfnamefont{A.}~\bibnamefont{Kleppe}}, \bibnamefont{and}
	\bibinfo{author}{\bibfnamefont{A.}~\bibnamefont{Jephcoat}},
	\bibinfo{journal}{Phys. Rev. Lett.} \textbf{\bibinfo{volume}{109}},
	\bibinfo{pages}{075501} (\bibinfo{year}{2012}).
	
	\bibitem{SI} See Supplemental Material at https://XXX for computational details, formation enthalpies, phase diagrams, electronic structures, phonon dispersion curves, superconducting properties, and structural information of alkline-earth borides, which includes Refs. [30, 31, 40, 46, 48-62]

	
	\bibitem[{\citenamefont{Hu et~al.}(2013)\citenamefont{Hu, Oganov, Zhu, Qian,
			Frapper, Lyakhov, and Zhou}}]{hu2013pressure}
	\bibinfo{author}{\bibfnamefont{C.~H.} \bibnamefont{Hu}},
	\bibinfo{author}{\bibfnamefont{A.~R.} \bibnamefont{Oganov}},
	\bibinfo{author}{\bibfnamefont{Q.}~\bibnamefont{Zhu}},
	\bibinfo{author}{\bibfnamefont{G.~R.} \bibnamefont{Qian}},
	\bibinfo{author}{\bibfnamefont{G.}~\bibnamefont{Frapper}},
	\bibinfo{author}{\bibfnamefont{A.~O.} \bibnamefont{Lyakhov}},
	\bibnamefont{and} \bibinfo{author}{\bibfnamefont{H.~Y.} \bibnamefont{Zhou}},
	\bibinfo{journal}{Phys. Rev. Lett.} \textbf{\bibinfo{volume}{110}},
	\bibinfo{pages}{165504} (\bibinfo{year}{2013}).
	
	\bibitem[{\citenamefont{Peng et~al.}(2012)\citenamefont{Peng, Miao, Wang, Li,
			and Ma}}]{peng2012predicted}
	\bibinfo{author}{\bibfnamefont{F.}~\bibnamefont{Peng}},
	\bibinfo{author}{\bibfnamefont{M.}~\bibnamefont{Miao}},
	\bibinfo{author}{\bibfnamefont{H.}~\bibnamefont{Wang}},
	\bibinfo{author}{\bibfnamefont{Q.}~\bibnamefont{Li}}, \bibnamefont{and}
	\bibinfo{author}{\bibfnamefont{Y.}~\bibnamefont{Ma}}, \bibinfo{journal}{J.
		Am. Chem. Soc.} \textbf{\bibinfo{volume}{134}}, \bibinfo{pages}{18599}
	(\bibinfo{year}{2012}).
	
	\bibitem[{\citenamefont{Sheng et~al.}(2011)\citenamefont{Sheng, Yan, Ye, Zheng,
			and Su}}]{sheng2011t}
	\bibinfo{author}{\bibfnamefont{X.~L.} \bibnamefont{Sheng}},
	\bibinfo{author}{\bibfnamefont{Q.~B.} \bibnamefont{Yan}},
	\bibinfo{author}{\bibfnamefont{F.}~\bibnamefont{Ye}},
	\bibinfo{author}{\bibfnamefont{Q.~R.} \bibnamefont{Zheng}}, \bibnamefont{and}
	\bibinfo{author}{\bibfnamefont{G.}~\bibnamefont{Su}}, \bibinfo{journal}{Phys.
		Rev. Lett.} \textbf{\bibinfo{volume}{106}}, \bibinfo{pages}{155703}
	(\bibinfo{year}{2011}).
	
	\bibitem[{\citenamefont{Zhang et~al.}(2017)\citenamefont{Zhang, Wang, Zhu, Pan,
			Han, Li, Zhao, Ma, Wang, Su et~al.}}]{zhang2017pseudo}
	\bibinfo{author}{\bibfnamefont{J.}~\bibnamefont{Zhang}},
	\bibinfo{author}{\bibfnamefont{R.}~\bibnamefont{Wang}},
	\bibinfo{author}{\bibfnamefont{X.}~\bibnamefont{Zhu}},
	\bibinfo{author}{\bibfnamefont{A.}~\bibnamefont{Pan}},
	\bibinfo{author}{\bibfnamefont{C.}~\bibnamefont{Han}},
	\bibinfo{author}{\bibfnamefont{X.}~\bibnamefont{Li}},
	\bibinfo{author}{\bibfnamefont{D.}~\bibnamefont{Zhao}},
	\bibinfo{author}{\bibfnamefont{C.}~\bibnamefont{Ma}},
	\bibinfo{author}{\bibfnamefont{W.}~\bibnamefont{Wang}},
	\bibinfo{author}{\bibfnamefont{H.}~\bibnamefont{Su}}, \bibnamefont{et~al.},
	\bibinfo{journal}{Nat. Commun.} \textbf{\bibinfo{volume}{8}},
	\bibinfo{pages}{683} (\bibinfo{year}{2017}).
	
	\bibitem[{\citenamefont{Eliashberg}(1960)}]{eliashberg1960interactions}
	\bibinfo{author}{\bibfnamefont{G.~M.} \bibnamefont{Eliashberg}},
	\bibinfo{journal}{Sov. Phys. JETP} \textbf{\bibinfo{volume}{11}},
	\bibinfo{pages}{696} (\bibinfo{year}{1960}).
	
	\bibitem[{\citenamefont{Sanna et~al.}(2018)\citenamefont{Sanna, Flores-Livas,
			Davydov, Profeta, Dewhurst, Sharma, and Gross}}]{sanna2018ab}
	\bibinfo{author}{\bibfnamefont{A.}~\bibnamefont{Sanna}},
	\bibinfo{author}{\bibfnamefont{J.~A.} \bibnamefont{Flores-Livas}},
	\bibinfo{author}{\bibfnamefont{A.}~\bibnamefont{Davydov}},
	\bibinfo{author}{\bibfnamefont{G.}~\bibnamefont{Profeta}},
	\bibinfo{author}{\bibfnamefont{K.}~\bibnamefont{Dewhurst}},
	\bibinfo{author}{\bibfnamefont{S.}~\bibnamefont{Sharma}}, \bibnamefont{and}
	\bibinfo{author}{\bibfnamefont{E.}~\bibnamefont{Gross}}, \bibinfo{journal}{J.
		Phys. Soc. Japan} \textbf{\bibinfo{volume}{87}}, \bibinfo{pages}{041012}
	(\bibinfo{year}{2018}).
	
	\bibitem[{\citenamefont{Giustino et~al.}(2007)\citenamefont{Giustino, Cohen,
			and Louie}}]{giustino2007electron}
	\bibinfo{author}{\bibfnamefont{F.}~\bibnamefont{Giustino}},
	\bibinfo{author}{\bibfnamefont{M.~L.} \bibnamefont{Cohen}}, \bibnamefont{and}
	\bibinfo{author}{\bibfnamefont{S.~G.} \bibnamefont{Louie}},
	\bibinfo{journal}{Phys. Rev. B} \textbf{\bibinfo{volume}{76}},
	\bibinfo{pages}{165108} (\bibinfo{year}{2007}).
	
	\bibitem[{\citenamefont{Liu et~al.}(2001)\citenamefont{Liu, Mazin, and
			Kortus}}]{liu2001beyond}
	\bibinfo{author}{\bibfnamefont{A.~Y.} \bibnamefont{Liu}},
	\bibinfo{author}{\bibfnamefont{I.}~\bibnamefont{Mazin}}, \bibnamefont{and}
	\bibinfo{author}{\bibfnamefont{J.}~\bibnamefont{Kortus}},
	\bibinfo{journal}{Phys. Rev. Lett.} \textbf{\bibinfo{volume}{87}},
	\bibinfo{pages}{087005} (\bibinfo{year}{2001}).
	
	\bibitem[{\citenamefont{Yildirim et~al.}(2001)\citenamefont{Yildirim,
			G{\"u}lseren, Lynn, Brown, Udovic, Huang, Rogado, Regan, Hayward, Slusky
			et~al.}}]{yildirim2001giant}
	\bibinfo{author}{\bibfnamefont{T.}~\bibnamefont{Yildirim}},
	\bibinfo{author}{\bibfnamefont{O.}~\bibnamefont{G{\"u}lseren}},
	\bibinfo{author}{\bibfnamefont{J.}~\bibnamefont{Lynn}},
	\bibinfo{author}{\bibfnamefont{C.}~\bibnamefont{Brown}},
	\bibinfo{author}{\bibfnamefont{T.}~\bibnamefont{Udovic}},
	\bibinfo{author}{\bibfnamefont{Q.}~\bibnamefont{Huang}},
	\bibinfo{author}{\bibfnamefont{N.}~\bibnamefont{Rogado}},
	\bibinfo{author}{\bibfnamefont{K.}~\bibnamefont{Regan}},
	\bibinfo{author}{\bibfnamefont{M.}~\bibnamefont{Hayward}},
	\bibinfo{author}{\bibfnamefont{J.}~\bibnamefont{Slusky}},
	\bibnamefont{et~al.}, \bibinfo{journal}{Phys. Rev. Lett.}
	\textbf{\bibinfo{volume}{87}}, \bibinfo{pages}{037001}
	(\bibinfo{year}{2001}).
	
	\bibitem[{\citenamefont{Lazzeri et~al.}(2003)\citenamefont{Lazzeri, Calandra,
			and Mauri}}]{lazzeri2003anharmonic}
	\bibinfo{author}{\bibfnamefont{M.}~\bibnamefont{Lazzeri}},
	\bibinfo{author}{\bibfnamefont{M.}~\bibnamefont{Calandra}}, \bibnamefont{and}
	\bibinfo{author}{\bibfnamefont{F.}~\bibnamefont{Mauri}},
	\bibinfo{journal}{Phys. Rev. B} \textbf{\bibinfo{volume}{68}},
	\bibinfo{pages}{220509} (\bibinfo{year}{2003}).
	
	\bibitem[{\citenamefont{Errea et~al.}(2014)\citenamefont{Errea, Calandra, and
			Mauri}}]{errea2014anharmonic}
	\bibinfo{author}{\bibfnamefont{I.}~\bibnamefont{Errea}},
	\bibinfo{author}{\bibfnamefont{M.}~\bibnamefont{Calandra}}, \bibnamefont{and}
	\bibinfo{author}{\bibfnamefont{F.}~\bibnamefont{Mauri}},
	\bibinfo{journal}{Phys. Rev. B} \textbf{\bibinfo{volume}{89}},
	\bibinfo{pages}{064302} (\bibinfo{year}{2014}).
	
	\bibitem[{\citenamefont{Friedrich et~al.}(2011)\citenamefont{Friedrich,
			Winkler, Juarez-Arellano, and Bayarjargal}}]{friedrich2011synthesis}
	\bibinfo{author}{\bibfnamefont{A.}~\bibnamefont{Friedrich}},
	\bibinfo{author}{\bibfnamefont{B.}~\bibnamefont{Winkler}},
	\bibinfo{author}{\bibfnamefont{E.~A.} \bibnamefont{Juarez-Arellano}},
	\bibnamefont{and}
	\bibinfo{author}{\bibfnamefont{L.}~\bibnamefont{Bayarjargal}},
	\bibinfo{journal}{Materials} \textbf{\bibinfo{volume}{4}},
	\bibinfo{pages}{1648} (\bibinfo{year}{2011}).
	
	\bibitem[{\citenamefont{Ishii et~al.}(2019)\citenamefont{Ishii, Liu, and
			Katsura}}]{ishii2019breakthrough}
	\bibinfo{author}{\bibfnamefont{T.}~\bibnamefont{Ishii}},
	\bibinfo{author}{\bibfnamefont{Z.}~\bibnamefont{Liu}}, \bibnamefont{and}
	\bibinfo{author}{\bibfnamefont{T.}~\bibnamefont{Katsura}},
	\bibinfo{journal}{Engineering} \textbf{\bibinfo{volume}{5}},
	\bibinfo{pages}{434} (\bibinfo{year}{2019}).
	
	\bibitem{SI3} Y. Wang and J. P. Perdew, Phys. Rev. B \textbf{43}, 8911 (1991).
	\bibitem{SI4} G. Kresse and J. Furthmüller, Phys. Rev. B \textbf{54}, 11169 (1996).
	\bibitem{SI5}  J. P. Perdew and Y. Wang, Phys. Rev. B \textbf{45}, 13244 (1992).
	\bibitem{SI6} P. E. Blöchl, Phys. Rev. B \textbf{50}, 17953 (1994).
	\bibitem{SI7} H. J. Monkhorst and J. D. Pack, Phys. Rev. B \textbf{13}, 5188 (1976).
	\bibitem{SI8} A. Togo, F. Oba, and I. Tanaka, Phys. Rev. B \textbf{78}, 134106 (2008).
	\bibitem{SI9} R. Dronskowski and P. E. Bloechl, J. Phys. Chem. \textbf{97}, 8617 (1993).
	\bibitem{SI10}V. L. Deringer, A. L. Tchougréeff, and R. Dronskowski, J. Phys. Chem. A \textbf{115}, 5461 (2011).
	\bibitem{SI11}S. Maintz, V. L. Deringer, A. L. Tchougréeff, and R. Dronskowski, J. Comput. Chem. \textbf{37}, 1030 (2016).
	\bibitem{SI12}P. Giannozzi, S. Baroni, N. Bonini, et al., J. Phys. Condens. Mat. \textbf{21}, 395502 (2009).
	\bibitem{SI13}G. Pizzi, V. Vitale, R. Arita, et al., J. Phys. Condens. Mat. \textbf{32}, 165902 (2020).
	\bibitem{SI14}S. Poncé, E. R. Margine, C. Verdi, and F. Giustino, Comput. Phys. Commun. \textbf{209}, 116 (2016)
	
	\bibitem{SI16} H. J. Choi, M. L. Cohen, and S. G. Louie, Phys. C (Amsterdam, Neth.) \textbf{385}, 66 (2003).
	\bibitem{SI17} E. R. Margine and F. Giustino, Phys. Rev. B \textbf{87}, 024505 (2013).
	
	\bibitem{SI19} S. D. Cataldo, W. V. D. Linden, L. Boeri, Phys. Rev. B, \textbf{102}, 014516 (2020).
	
\end{thebibliography}
\end{document}